\documentclass[fleqn,usenatbib]{mnras}
\usepackage{newtxtext,newtxmath}

\usepackage[T1]{fontenc}
\usepackage{CJK}

\DeclareRobustCommand{\VAN}[3]{#2}
\let\VANthebibliography\thebibliography
\def\thebibliography{\DeclareRobustCommand{\VAN}[3]{##3}\VANthebibliography}

\usepackage{graphicx}	
\usepackage{amsmath}	
\usepackage{orcidlink}

\newcommand{\gray}{$\gamma$-ray}
\newcommand{\enc}{$\Gamma$}
\newcommand{\lumg}{$L_\gamma$}
\newcommand{\nmsp}{$N_{\rm MSP}$}

\title[Gamma-rays from Globular Clusters]{A Gamma-ray Stacking Survey of Fermi-LAT Undetected Globular Clusters}

\author[Henry et al.]{
Owen K. Henry\orcidlink{0000-0002-0472-9195},$^{1,2}$\thanks{E-mail: ohenry@gradcenter.cuny.edu}
Timothy A.D. Paglione\orcidlink{0000-0001-9139-0945},$^{3,2,1}$
Yuzhe Song(宋宇哲)\orcidlink{0000-0002-2080-9232},$^{4,5}$
Joshua Tan\orcidlink{0000-0003-3835-8115},$^{6,2}$
\newauthor David Zurek,$^2$
and Vanessa Pinto$^{7}$
\\
$^{1}$Department of Physics, The Graduate Center, City University of New York, 365 Fifth Ave., New York, NY 10016, USA\\
$^{2}$Department of Astrophysics, American Museum of Natural History, Central Park West at 79th Street, New York, NY 10024, USA\\
$^{3}$Department of Earth \& Physical Sciences, York
College, City University of New York, 94-20 Guy R. Brewer Blvd., Jamaica, NY 11451, USA\\
$^{4}$Center for Astrophysics and Supercomputing, Swinburne University of Technology, John St, Hawthorn, VIC 3122, Australia\\
$^{5}$OzGrav, ARC Center for Excellence in Gravitational Wave Discovery\\
$^{6}$Department of Natural Sciences, LaGuardia Community College, City University of New York, 31-10 Thomson Ave, Long Island City, NY 11101\\
$^{7}$Max Planck Institute for Radio Astronomy, Auf dem H{\"u}gel 69, 53121, Bonn, Germany\\}

\date{Accepted 2024 October 15. Received 2024 September 24; in original form 2024 June 13}

\pubyear{2024}

\begin{document}
\begin{CJK*}{UTF8}{gbsn}
\label{firstpage}
\pagerange{\pageref{firstpage}--\pageref{lastpage}}
\maketitle

\begin{abstract}
We present evidence for \gray\ emission from a stacked population of 39 high-latitude globular clusters (GCs) not detected in the Fermi Point Source Catalog, likely attributable to populations of millisecond pulsars within them. In this work, we use 13 years of data collected by the Large Area Telescope aboard the Fermi Gamma-Ray Space Telescope to search for a cumulative signal from undetected GCs and compared them to control fields (CFs), selected to match the celestial distribution of the target clusters so as to distinguish the \gray\ signal from background emission. The joint likelihood distribution of the GCs has a significant separation ($\sim4\sigma$) from that of the CFs. We also investigate correlations between detected cluster luminosities and other cluster properties such as distance, the number of millisecond pulsars associated with each cluster, and stellar encounter rate but find no significant relationships.
\end{abstract}




\section{Introduction} \label{sec:intro}

\par The Large Area Telescope (LAT) aboard the Fermi Gamma-ray Space Telescope has been measuring the most energetic phenomena in the universe since 2008. During its mission, it has detected 
\gray s from many different source classes including globular clusters (GCs). The first GC detected by the LAT was 47 Tuc 
\citep{2009Sci...325..845A}, and soon after there were studies of \gray\ emission from other GCs using the LAT such as Terzan 5 \citep{2010ApJ...712L..36K}, M15 \citep{2016MNRAS.459...99Z}, M80 \citep{2011ApJ...729...90T} and many others \citep{2018MNRAS.480.4782L, 2016JCAP...08..018H, 2022RAA....22k5013Y, 2022RAA....22e5019Y}. Today, there are a total of 32 detected globular clusters in the 12-year LAT catalog, 4FGL-DR3  
\citep[][hereafter 4FGL]{2022yCat.9067....0A}. GCs have proven to be an ideal environment for millisecond pulsars (MSPs) because MSPs are most likely formed through recycling processes in binary systems. Thus, the high density and encounter rate of a GC can foster efficient MSP formation \citep{1991PhR...203....1B, 2006csxs.book..623T, 2020arXiv201111385D}. When the neutron star's companion overflows its Roche lobe, material accretes onto the neutron star depositing angular momentum and decreasing the neutron star spin period down to the millisecond regime. 
Such mass-exchange binary systems appear as low-mass X-ray binaries (LMXBs) \citep{bhattacharya_1996} and are the prime progenitor candidates of MSPs \citep{1982Natur.300..728A}. Per unit mass, LMXBs are two orders of magnitude more abundant in GCs than in the Galactic field \citep{1988Natur.336...48G, 1975ApJ...199L.143C, 1975Natur.253..698K}. MSPs are found in excess in GCs at a similar order of magnitude. 
To date, over 330 MSPs have been detected in at least 44 GCs \citep{pfreire}.

The hypothesis that MSPs are the primary source of \gray s from GCs is supported by detections of pulsed \gray\ emission in the millisecond regime from GCs over the mission time of Fermi \citep{2011Sci...334.1107F, 2013ApJ...778..106J, 2023ApJ...945...70Z}. To this end, \citet{2022ApJ...927..117W} investigated spectral energy distributions (SEDs) of 104 MSPs detected with LAT and compared them to SEDs 
of detected GCs in the 4FGL. They aimed to identify contributions from two leptonic processes that are thought to govern the emission physics of \gray s around MSPs: curvature radiation coming directly from the pulsars, 
and inverse Compton (IC) scattered background photons from 
the CMB, the Galactic radiation field, or the dense radiation field of the cluster itself \citep{2005ApJ...622..531H}. \citet{2022ApJ...927..117W} concluded that it is unclear which emission mechanism dominates. 

In this study, we look for \gray\ signals from GCs yet undetected by Fermi. We also aim to leverage the low luminosity clusters examined in this study to extend correlation analyses between the \gray\ luminosity (\lumg) and various cluster properties. In addition, this correlation studies could help inform follow-up observations of clusters to search for undetected radio pulsars. In this study, we conduct correlation tests similar to that of \citet{2019MNRAS.486..851D}, \citet{2021MNRAS.507.5161S}, and \citet{2023arXiv231015859F} between the \gray\ luminosity (\lumg\ ) of detected GCs and three physical properties that are related to the dynamics of the GC: the stellar encounter rate, the number of MSPs, and the photon field density.

We present this work as follows: In \S \ref{sec:obs} the data selection criteria for the \gray\ target GCs are described and their data processing procedure from \texttt{fermipy} is discussed. In \S \ref{sec:results} we describe our stacking procedures. In \S \ref{sec:disc} the data analysis and results are discussed, 
including stack significance and correlation analyses. In \S \ref{sec:5} we review and summarize our results as well as 
suggest possible directions forward.

\section{Observations}\label{sec:obs}

\subsection{Target Selection}\label{sec:targs}

We select our set of target GCs from \citet{2010arXiv1012.3224H} with a Galactic latitude cut of $|b| > 20^{\circ}$ to avoid the complex background of the \gray -bright plane of the Galaxy and excluding clusters that are already detected in the 4FGL. These selection criteria yield 39 target GCs. 

\begin{figure*}
    \centering
    \includegraphics[width=\textwidth]{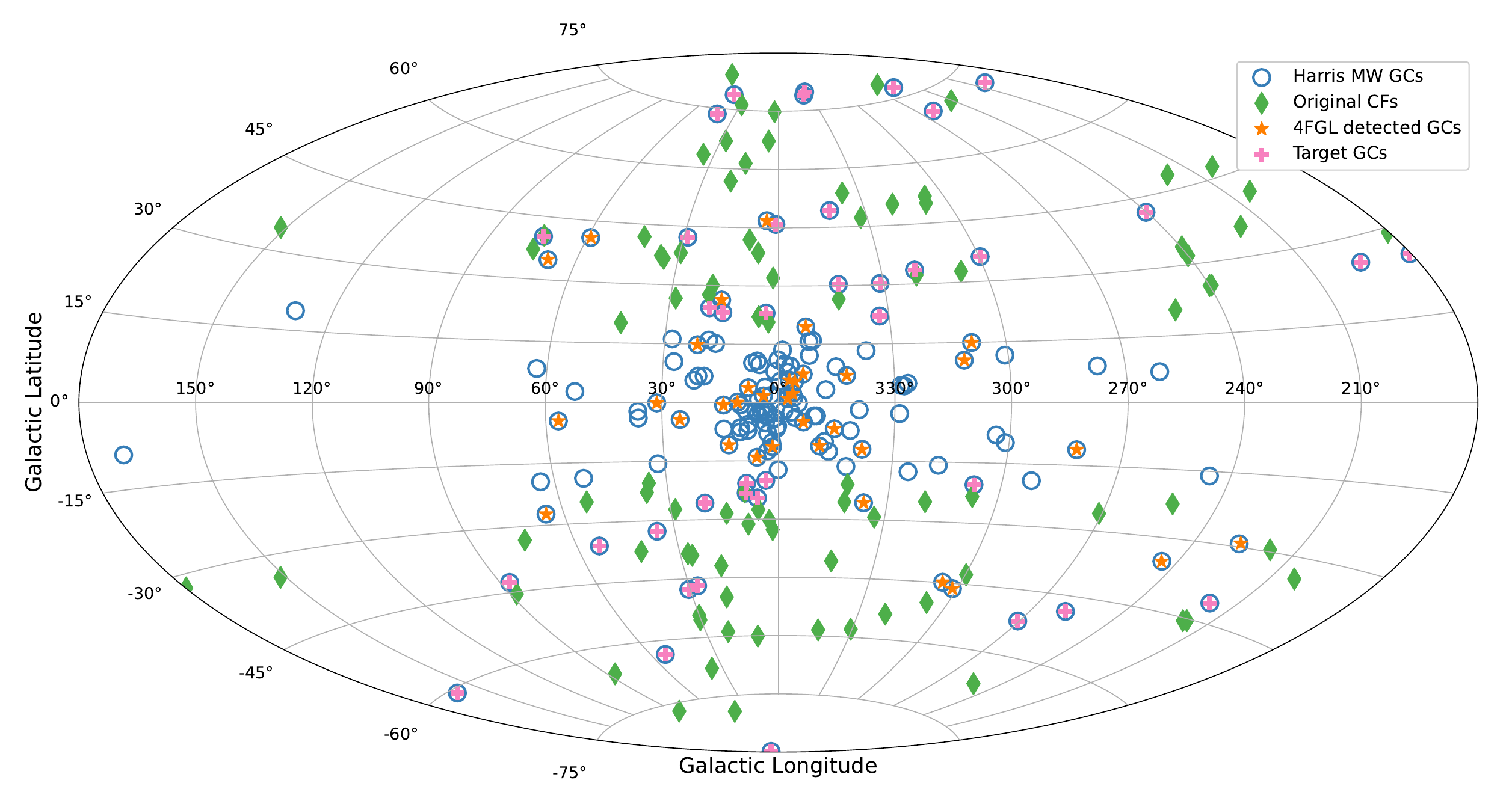}
    \caption{All-sky map of the target clusters analyzed in this work (crosses), the detected GCs in the 4FGL (stars), and the control field test sources (diamonds).}
    \label{fig:allsky}
\end{figure*}

For comparison, we select control fields (CFs) by generating a randomly distributed sample matching the Galactic latitude and 
longitude distributions of the target GCs, and $|b| > 20^{\circ}$. 
To avoid contamination, we excluded CFs centered within $1.7^{\circ}$, a distance of roughly twice the containment radius, of our targets or 4FGL sources. In the end, we use 90 CFs for the analysis. With roughly double the CF test sources as target GCs, we sufficiently capture the Poisson variance 
while minimizing computational expense. This procedure for selecting CFs is standard practice  \citep[e.g., ][]{2024PhRvD.109f3024M, 2019ApJ...882L...3P, 2023PhRvD.107h3030D}. The locations of all target sources, control fields, and 4FGL-detected clusters are shown in Fig.~\ref{fig:allsky}. We test the validity of this population selection of CF test sources in \S \ref{sec:toy}. 

\subsection{Binned Likelihood Analysis}\label{sec:like}
We adopt a typical maximum likelihood analysis to search for \gray\ emission from our targets \citep{1996ApJ...461..396M}. Thirteen years of LAT data between mission elapsed time (MET) 239160000 s and 651715205 s were used in this study. We filtered data using a zenith angle cut of $90^{\circ}$ to avoid contamination from the Earth. The photon energy range for analysis is 300 MeV to 100 GeV, 
which is split into 30-logarithmically spaced bins\footnote{Fermi science tools and fermipy tutorials: https: //fermi.gsfc.nasa.gov/ssc/data/analysis/scitools/ and http://fermipy.readthedocs.io/en/latest/quickstart.html}. This energy range has been shown to maximize the sensitivity of the analysis \citep{2023MNRAS.524.5854S, 2020ApJ...897..177P}. All-sky livetime and exposure cubes were created for all 129 (39 targets plus 90 CF) regions of interest (ROIs) that we consider in this analysis. We use the third revision of the Pass 8 (P8R3) instrument response function (P8R3\_SOURCE\_V3), the most recent Galactic emission model (\texttt{gll\_iem\_v07.fits}), and isotropic background emission model (\texttt{iso\_P8R3\_SOURCE\_V3\_v1.txt}) \citep{2009ApJ...703.1249A} with the default event class and type (evclass = 128, evtype = 3)\footnote{https://fermi.gsfc.nasa.gov/ssc/data/access/lat/BackgroundModels.html}.
 
We perform this analysis using \texttt{fermipy}, a Python package that facilitates analysis of LAT data with the Fermi Science Tools within the open source distribution of Python, Anaconda \citep{2019ascl.soft05011F, 2017ICRC...35..824W, anaconda}. We perform the maximum likelihood test for the presence of a \gray\ point source at each target's location on the sky.  The result of the likelihood analysis is the Test Statistic (TS), defined as TS $ = 2\text{ln}(L/L_0)$, where $L$ is the likelihood of a point source being present at the center of the ROI, and $L_0$ is the null hypothesis that there is no central source \citep{1996ApJ...461..396M}. The detection significance can be estimated from $\sqrt{\textrm{TS}}$, and we adopt the usual detection threshold of TS $>25$ \citep{2022yCat.9067....0A}.

The ROIs are $21^{\circ}\times 21^{\circ}$ square cutouts on the sky centered around each target coordinate. We model an additional point source at the center of the ROI with a spectral model that is described below in \S \ref{sec:model}.  Spectral parameters of 4FGL sources within $5^{\circ}$ of the center of the ROI are free to be fit, and those outside remain fixed. The spectral model adopted in this study is discussed in the following section (\S \ref{sec:model}).

We search for additional unmodeled point sources by generating TS maps for each ROI using \texttt{gta.find\_sources}. We search for power law sources with a spectral index of $-2$ outside of a $0.3^{\circ}$ radius from the ROI center and then identify sources with a minimum detection threshold of TS $>25$. Sources that peak above this threshold have their spectral parameters fit and are then 
added to the model. TS values for all target GCs in this study are presented in Table~\ref{tab:table}. We conduct an identical analysis for the CF test sources.

\subsection{Spectral Modeling of Globular Clusters}\label{sec:model}

We test two different spectral models to maximize the sensitivity of our analysis. The spectral models most commonly used for GCs are \texttt{LogParabola} (LP) and \texttt{PowerLawSuperExpCutoff} (PLEC) \footnote{https://fermi.gsfc.nasa.gov/ssc/data/analysis/scitools/source\_models.html}. The Fermi LAT consortium typically uses the LP spectral model to fit GCs, but several studies fit GC spectra with the PLEC model \citep{2019MNRAS.486..851D, 10.1111/j.1365-2966.2007.11664.x, 2018MNRAS.480.4782L}. The spectral flux given by the PLEC model is

\begin{equation}
    \frac{dN}{dE} = N_0(\frac{E}{E_0})^{\gamma}e^{-({E}/{E_c})^b}
\end{equation}

\noindent
where $N_0$ is the normalization prefactor, $\gamma$ is the power law spectral index, $E_0$ is the energy scale factor, $E_c$ is the cutoff energy, and $b$ is a second power law index that determines the curvature at the cutoff. We also test the LP model:  
\begin{equation}
   \frac{dN}{dE} = N_0(\frac{E}{E_0})^{-(\gamma+\beta\text{log}({E}/{E_0}))}
\end{equation}

\noindent 
where $\beta$ measures the spectral curvature 
\citep{2006A&A...448..861M}. Again $E_0$ is a fixed scale parameter, $\gamma$ is the spectral index, and $N_0$ is the normalization.  We find no significant difference in TS when modeling the 4FGL GCs with the PLEC model compared to that of an LP model. So, we adopt the PLEC spectral model fits for all subsequent analyses and discussions in this work. The principal advantage of using the PLEC model is that it has fewer degrees of freedom, tending to yield a higher significance for a given TS. The cutoff energy ($E_c$) and the energy scale ($E_0$) were fixed at 1000 MeV. The second power law index was also fixed at $b = 1$ (for justification, see \citet{2023MNRAS.524.5854S}). Only the spectral index 
and normalization prefactor is free to fit. 

\section{Results}\label{sec:results}

\subsection{Cumulative TS Distributions}\label{sec:sigTS}
\begin{figure*}
    \centering
    \includegraphics[width = \textwidth]{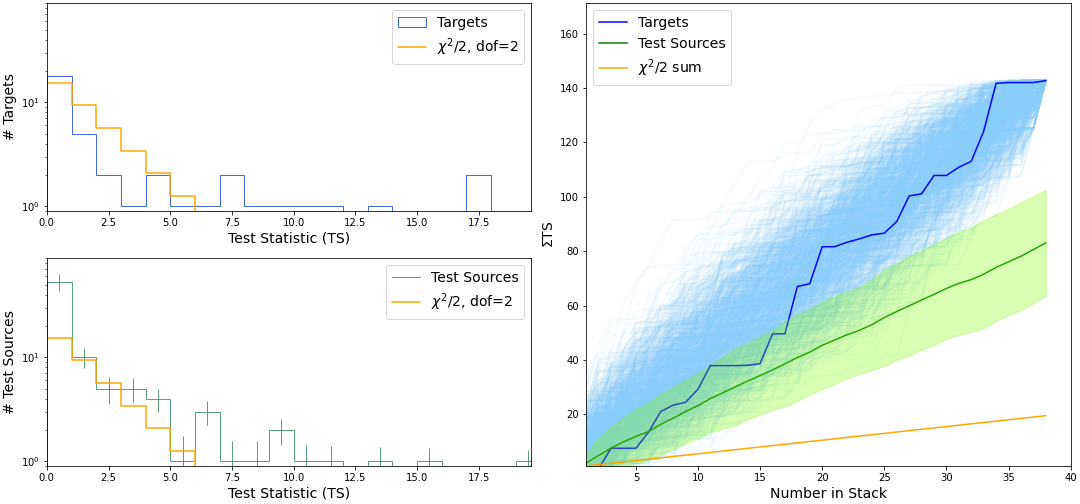}
    \caption{
    (Left) Histograms of the TS values of target GCs (top) and CF test sources (bottom). The CF test source histogram and error bars are the average and standard deviation of randomly selecting 39 of the 90 sources 100 times. The theoretical null ($\chi^2/2$) is shown for comparison in orange.
    (Right) Cumulative TS of the GCs (blue) compared to CF test sources (green).
    The target TS stack is a randomly ordered sum of all measured TS values (Table \ref{tab:table}). The light blue envelope shows 1000 iterations of the sum done in different random orders. 
    The CF test source stack is the sum of 39 randomly sampled fields out of the 90 CF test sources; the green line and shaded envelope depict the mean sum and standard deviation, respectively, as a function of stacked ROIs. The sum of the theoretical null is also shown (orange).}
    \label{fig:cumuTS}
\end{figure*}

The target GCs and CFs are stacked following the procedure developed by \citet{2023MNRAS.524.5854S}, which was adapted from the technique of \citet{2012A&A...547A.102H}. Fig.~\ref{fig:cumuTS} shows the TS distributions of the central sources in the target and CF ROIs. The $\chi^2/2$ distribution, corresponding to the theoretical null \citep{d543aecb-cd73-36d5-9101-f08a74f8e8c6}, is also shown 
for comparison.
%
We sum the TS values of the 39 target GCs and compare the result to an equivalent cumulative TS distribution for the CF test sources (Fig.~\ref{fig:cumuTS}).  For the CF test source sum, we randomly draw 39 of the 90 fields 100 times and calculate the average sum of the cumulative TS values as a function of the stacked number of ROIs. The stack of the target clusters is displayed with 1000 random reorderings of the sum to illustrate its variation. There is a separation of $\Delta$TS = 59 between the target GC and CF test source populations. We quantify the separation significance in \S \ref{sec:toy}. Finally, both the target GC and CF test source stacks diverge significantly from the theoretical null (which also stacks to a non-zero cumulative TS) indicating the signal in both the sample and the blank sky. 
 
 \begin{table*}
    \centering
    \caption{Maximum Likelihood Results for Target Globular Clusters}
    \label{tab:table}
    \begin{tabular}{lccr|lccr}
    
    \hline
    Name & RA ($^{\circ}$) & Decl ($^{\circ}$) & TS & Name & RA ($^{\circ}$) & Decl ($^{\circ}$) & TS   \\
    \hline
    NGC 288 & 13.198 & -26.590 & $<0.1$ & Whiting & 30.506 & -3.248 & 0.711   \\
    NGC 1261 & 48.064 & -55.217 & $<0.1$ & AM1 & 58.761 & -49.614 & 3.032      \\
    Eridanus & 66.185 & -21.187 &  0.277 & NGC 2419 & 114.535 & 38.882 & $<0.1$\\
    Ko 2 & 119.567 & 26.246 & $<0.1$    & Pal 3 & 151.381 & 0.071 & 1.623     \\
    Pal 4 & 172.320 & 28.973 & 1.246    & Ko 1 & 179.828 & 12.253 & $<0.1$    \\
    NGC 4147 & 182.526 & 18.542 & 1.033 & NGC 4590 & 189.860 & -26.742 &$<0.1$\\
    NGC 5024 & 198.230 & 18.169 & 4.811 & NGC 5053 & 199.112 & 17.698 & 8.682 \\
    NGC 5272 & 205.546 & 28.375 & 7.699 & AM4 & 208.958 & -27.173 & 2.180     \\
    NGC 5466 & 211.363 & 28.534 & 0.123 & NGC 5634 & 217.405 & -5.976 & $<0.1$\\
    NGC 5694 & 219.902 & -26.538 & 7.519&  IC~4499 & 225.077 & -82.213 & 17.626\\
    NGC 5824 & 225.993 & -33.067 & $<0.1$ &  Pal 5 & 229.022 & -0.108 & 4.272 \\
    NGC 5897 & 229.352 & -21.010 & 0.601 &  Pal 14 & 242.770 & 14.958 & 1.456\\
    NGC 6171 & 248.133 & -13.053 & $<0.1$ & NGC 6229 & 251.745 & 47.527 & 2.366\\
    NGC 6254 & 254.287 & -4.099 & 5.802 & Pal 15 & 255.010 & 0.542 & 0.772 \\
    Terzan 7 & 289.432 & -34.657 & $<0.1$ &  Arp2 & 292.191 & -30.353 & 11.056\\ 
    NGC 6809 & 294.997 & -30.962 & 10.954 & Terzan 8 & 295.437 & -34.0002 & 6.740\\
    NGC 6864 & 301.520 & -21.921 & 1.025  &  NGC 6981 & 313.366 & -12.537 & $<0.1$\\
     NGC 7089 & 323.372 & -0.005 & 0.031 &   NGC 7099 & 325.091 & -23.179 & 17.409 \\
     Pal 12 & 326.661 & -21.251 & 9.426 &  Pal 13 & 346.685 & 12.772 & 0.635 \\
     NGC 7492 & 347.111 & -15.611 & 13.59\\ 
     \hline
       \end{tabular}

\end{table*}
\subsection{Parameter Space Stacking Analysis}\label{sec:param}
The target GCs and CF test sources undergo another TS stacking procedure by fitting their spectral properties 
similar to \citet{2019ApJ...882L...3P}. 

\begin{figure}
    \centering
    \includegraphics[width = \columnwidth]{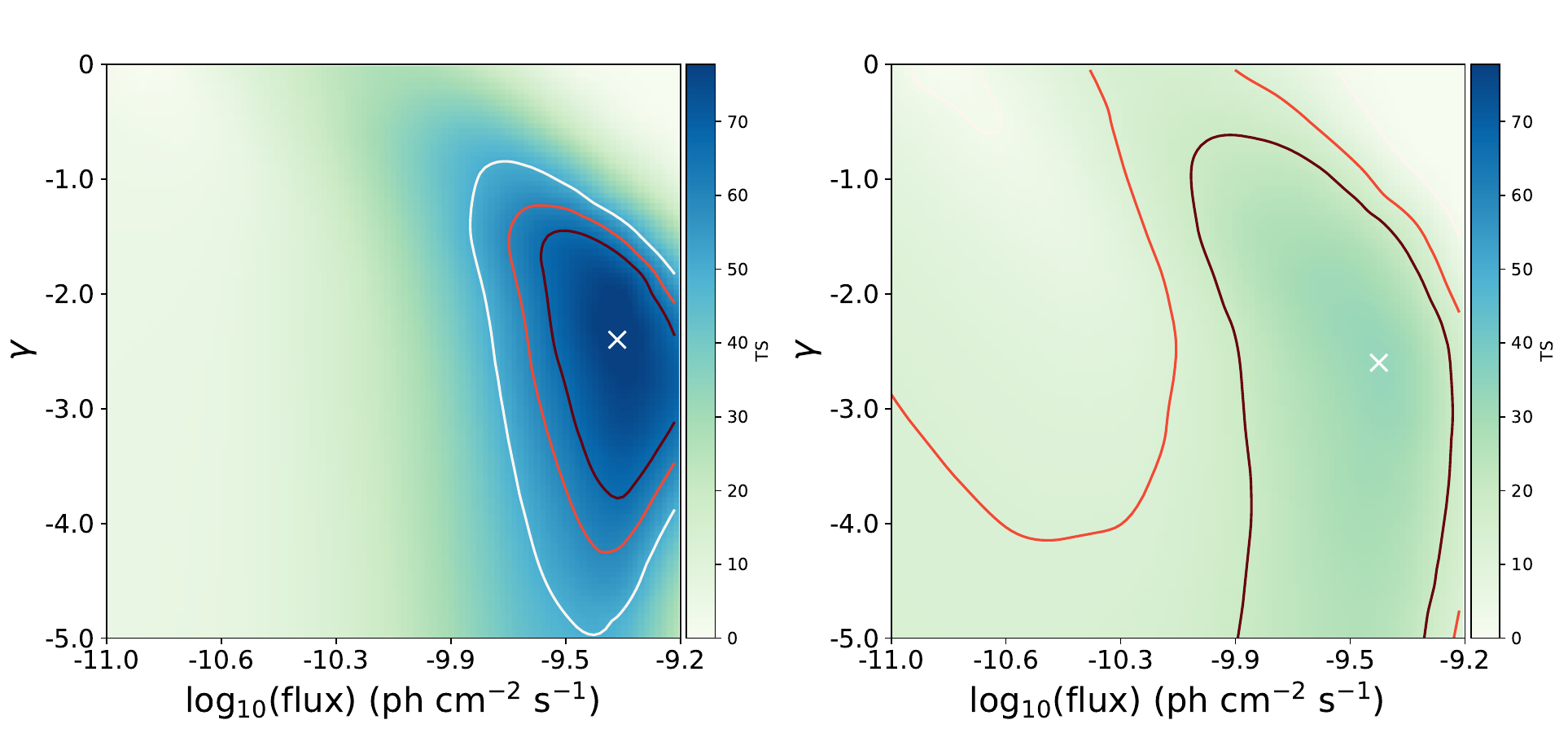}
    \caption{(Left) PLEC model parameter stack for the target GCs. The parameter stack peaks at a spectral index of 
    $-2.3^{+0.9}_{-1.5}$, log(flux) of $-9.2^{+0.2}_{-1.3}$ 
    ph cm$^{-2} \text{s}^{-1}$, and $\text{TS} = 79$. 
    (Right) CF test sources PLEC fit parameter stack with maximum $\text{TS} = 32$. 
    The contours represent the $3, 4, \: \text{and} \: 5\sigma$ distances from the best-fit location in the map. This map was created by again sampling 39 out of the 90 CF test sources 1000 times and then averaging over the 1000 random samplings. The color scale is set to the peak value of the target GC parameter space map.
    }
    \label{fig:param}
\end{figure}

As described in \S \ref{sec:like}, a point source with a PLEC spectrum is placed at the coordinates of the target GC. In this fit, however, only the normalization of the Galactic and diffuse background models are free to fit. We compute the log-likelihood for the ROI for a fixed spectral index and flux and repeat this process over a grid of $\gamma$ and flux values. 
To convert this log-likelihood map into a TS map, we adopt a null likelihood ($L_0$) at the lowest flux and index coordinate of the parameter space, subtract it from the rest of the map, and multiply by 2. The TS maps of each target GC are stacked to construct a parameter space significance map for our undetected cluster population. We take a resampled average of the CF test sources shown in the right panel of Fig.~\ref{fig:param}. A separation between the target GCs and CF test sources is again evident ($\Delta$TS = 47) between the peak 
TS of the targets and the controls. For the target GCs, the significance peaks at $\gamma = -2.7^{+0.8}_{-1.5}$ and log(flux) $=-9.2^{+0.2}_{-1.3}$ (ph cm$^{-2}$ s$^{-1}$). 

Two of the sources in our target GC population have a TS $> 16$: NGC~7099 (M~30) and IC~4499. We discuss these sources in Appendix \ref{appxA}. Even after removing these two sources from the cumulative TS and parameter space stacking analyses, the target GC population is still more significant than the CF test sources, which we quantify in \S\ref{sec:toy}. 

\section{Discussion}\label{sec:disc}
\subsection{Stack Detection Significance}\label{sec:toy}
To robustly quantify the separation between our target GC and CF cumulative TS (Fig.~\ref{fig:cumuTS}), and thus establish a stack detection, we 
model the 
\gray\ photon counts from a sub-threshold population.
The model 
assumes 
that the photon counts per pixel follow a Poisson distribution, the high latitude \gray\ background is isotropic, and the spatial distribution of source counts is a 2-D Gaussian with a FWHM of $0.5^{\circ}$. 
We assume a source population that follows a power law distribution with flux, $N \propto S^{-\alpha}$, where $N$ is the number of sources per bin, $S$ is the photon flux, and $\alpha$ is the power law index. The normalization depends on the signal-to-noise ratio (SNR). Here the SNR compares the total counts on source (source plus background) to the total background counts in the same number of pixels. The ratio of source to background counts, which we will refer to as the gain $g = \mathrm{SNR}-1$, is necessarily well below unity for a sub-threshold source. 
Our objective is to reproduce the observed cumulative TS values of the target GCs and CF test sources using a distribution of model point sources with varied $g$ and a power law index of $\alpha$.  
We create a set of models given the slope of the distribution ($\alpha$) and the size of the domain $0<g<g_{\rm max}$, where $g_{\rm max}$ corresponds to the maximum gain of the power law distribution 
as defined by the \texttt{scale} keyword argument in \texttt{scipy.stats.powerlaw.rvs}. 

\begin{figure}
    \centering
    \includegraphics[width = \columnwidth]{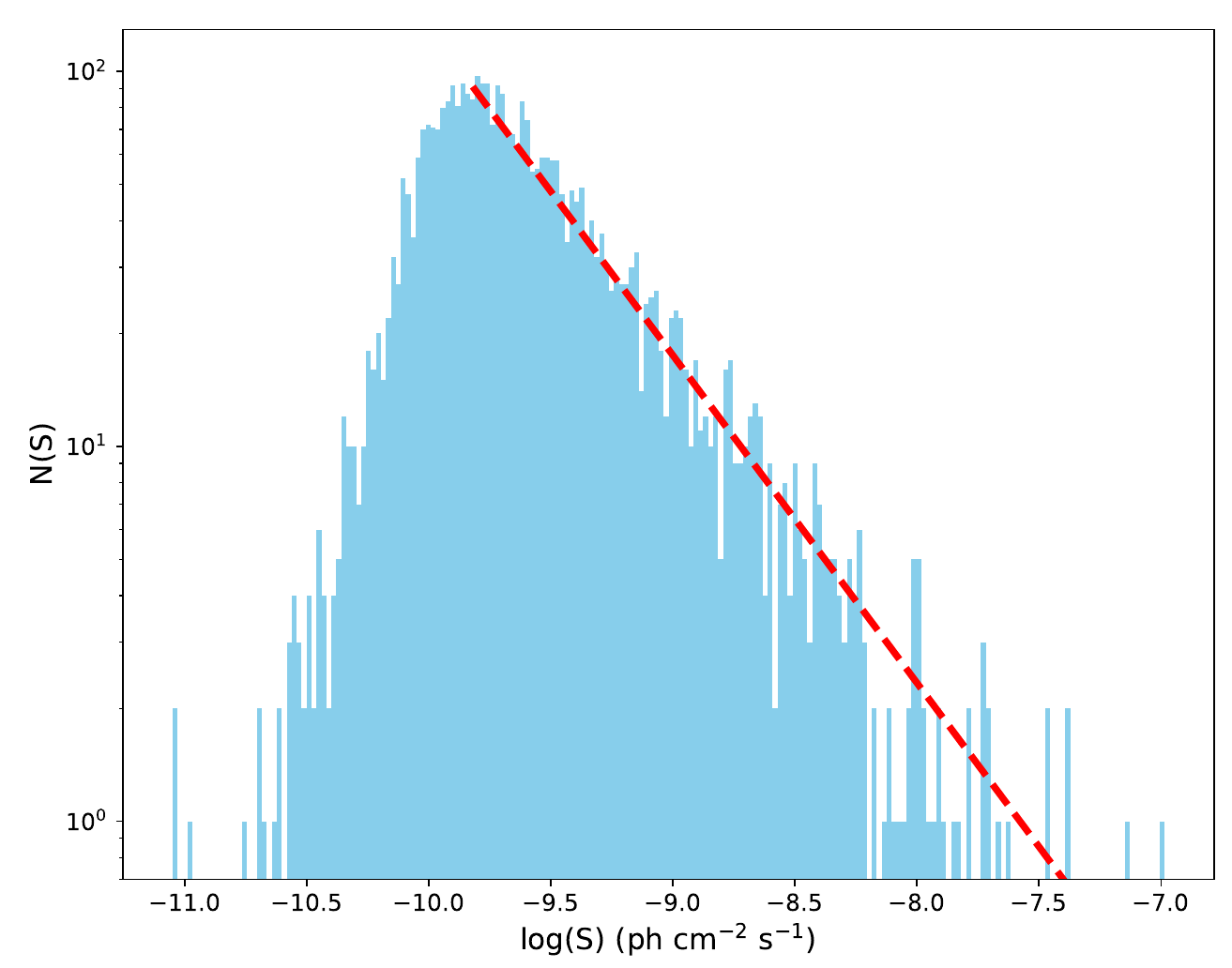}
    \caption{Histogram of high latitude 4FGL source fluxes. The linear fit (red dashed line) has a slope of $\alpha = -0.873$. The plotted fluxes are the \texttt{flux1000} measurements from the 4FGL-DR3 catalog (\texttt{gll\_psc\_v28.fit)}.}
    \label{fig:dNdS}
\end{figure}

We explore $-0.95 \leq \alpha \leq -0.75$, which is centered on a fit of the source count of high latitude 4FGL sources (Fig.~\ref{fig:dNdS}), measured to be $\alpha = -0.87 \pm 0.03$. We use 
$0.1 \leq g_{\rm max} \leq 0.6$, which yields cumulative TS values of the modeled stacks that encompass the observational results (Fig.~\ref{fig:cumuTS}).

\begin{figure}
    \centering
     \includegraphics[width = \columnwidth]{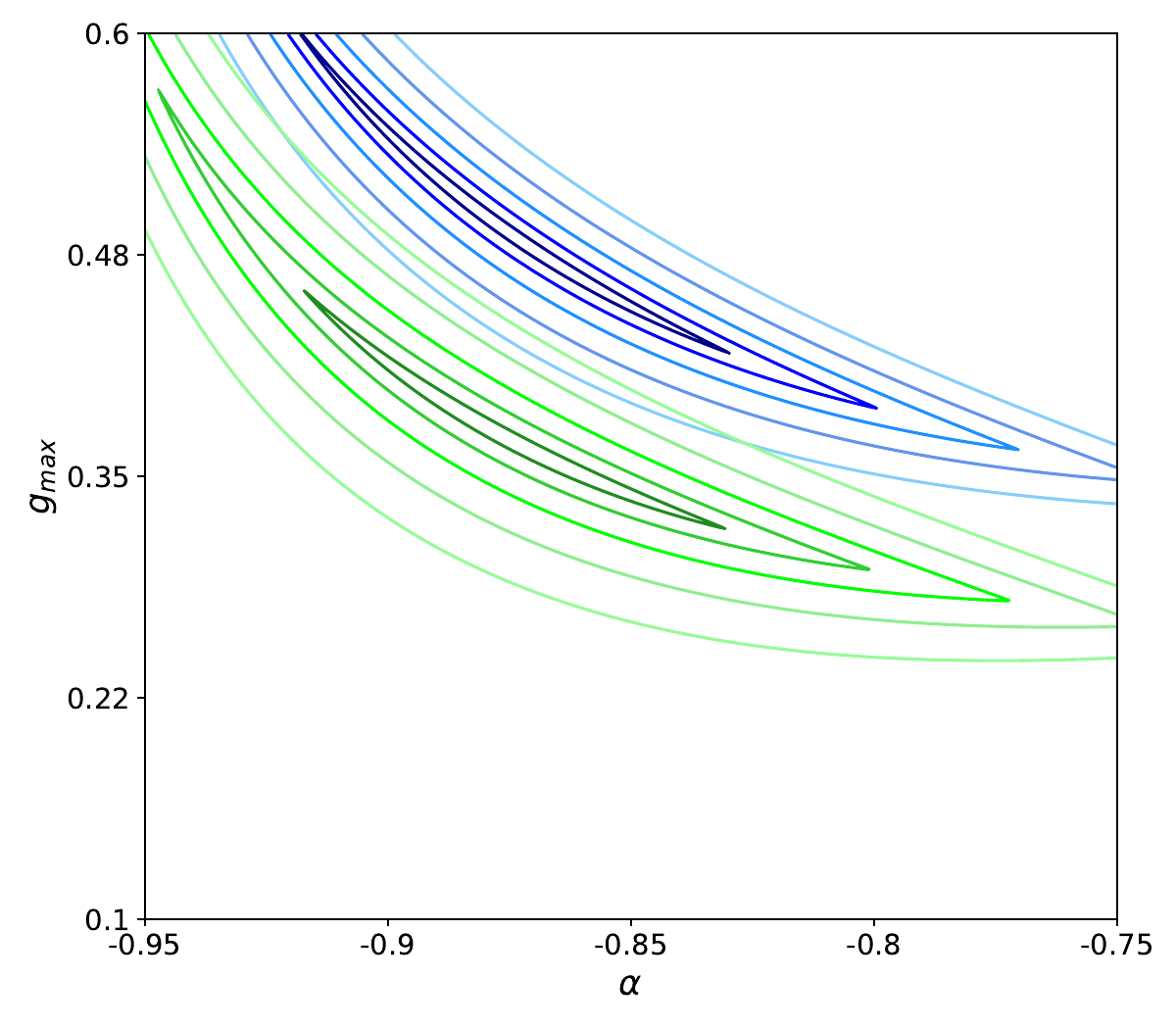}
    \caption{
    Significance maps ($\sqrt{\Delta \mathrm{TS}}$)
    characterizing the source populations underlying the GC (blue) and CF (green) stacks.  The contours are 1, 2, 3, 4, and 5 $\sigma$ 
    away from each global minimum.}
    \label{fig:bananas}
\end{figure}

For each value of $g_{\rm max}$ and $\alpha$, a model population of 10,000 sources is synthesized by building a distribution of $g$ values.  For each model source, the gain $g$ and the background counts are used to calculate the TS value \citep{1996ApJ...461..396M}. We adopt an estimate of 20 background counts per $0.1\degr$ pixel. From these 10,000 model sources, we randomly draw 39 to match the number of our target GCs. Their TS values are summed, returning a model cumulative TS. This model stacking is done 1000 times for a given  $g_{\rm max}$ and $\alpha$ and then averaged. Finally, we calculate the absolute value of the difference between the model results and the cumulative TS values of the target GCs and CFs (the maximum values of the distributions shown in the right panel of Fig.~\ref{fig:cumuTS}, or 144 and 85, respectively). The final target GC and CF significance distributions, $\sqrt{|\Delta\rm{TS}|}$, are shown in Fig.~\ref{fig:bananas}. We use the Kullback-Leibler divergence \citep{10.1214/aoms/1177729694} implemented in SciPy with \texttt{scipy.special.kl\_div}, to estimate the significance of the difference between these two 
distributions. According to that approach, the target GC stack is detected with a significance of $4.7 \sigma$ over the controls. Excluding NGC~7099 and IC~4499 reduces this significance to $3.5\sigma$, still indicating an excess signal from the target population over the controls. 

Given the sparseness of our GC target population, we test the appropriateness of our CFs by probing whether our target GCs are biased towards regions of excess or anomalous \gray\ background. We generate a new CF test source population consisting of 78 ROIs with centers $\pm 5\degr$ in Galactic longitude from each of our targets, again avoiding 4FGL sources.  
While we expect these test CFs to be biased due to a systematic contamination by the sub-threshold targets, it still may serve as a valuable test given the patchy and latitude-dependent structure of the LAT sensitivity \citep[][hereafter 3PC]{2023ApJ...958..191S}.
The latitude distributions of our targets, these CFs and the original CFs, are statistically indistinguishable. The 3PC LAT sensitivity distributions at their locations are also statistically equivalent.

Comparing the stack of these CFs to the target stack still yields a detection significance of $3.7 \sigma$. As anticipated, the cumulative TS value of the latitude-matched CF test sources is greater than the original CF test source population by $3.4 \sigma$.  
While the original CFs are likely therefore more appropriate, we conservatively report a $\sim4\sigma$ stack detection of the GC population in this study. 

\subsection{Correlation Analysis}\label{sec:corr}
We investigate possible correlations between the \gray\ luminosity (\lumg) and physical cluster parameters likely to be associated with the high energy emission from GCs, namely 
the number of MSPs (\nmsp) and stellar encounter rate (\enc). From \citet{2013ApJ...766..136B} \enc\ is taken to be: 
 
\begin{equation}
     \Gamma = A \frac{4 \pi}{\sigma_c}\int \rho(r)^2 r^2 dr ,
\end{equation}
where $\sigma_c$ is the velocity dispersion at the core radius, and $\rho(r)$ is the stellar density profile of the cluster. The line-of-sight integration is performed out to the half-light radius. As defined, \enc\ is an index that measures the average rate of encounters within a GC. The constant $A$ is such that \enc\ is normalized to 1000 encounters in the cluster 47 Tucanae \citep{2013ApJ...766..136B}. 
LMXBs, thought to be the progenitors of MSPs, are overabundant in GCs due to the formation of these systems through stellar interactions. It follows that for GCs, \enc\ could be a tracer of MSPs and thus \gray s. 

\begin{figure*}
    \centering
     \includegraphics[width = \textwidth]{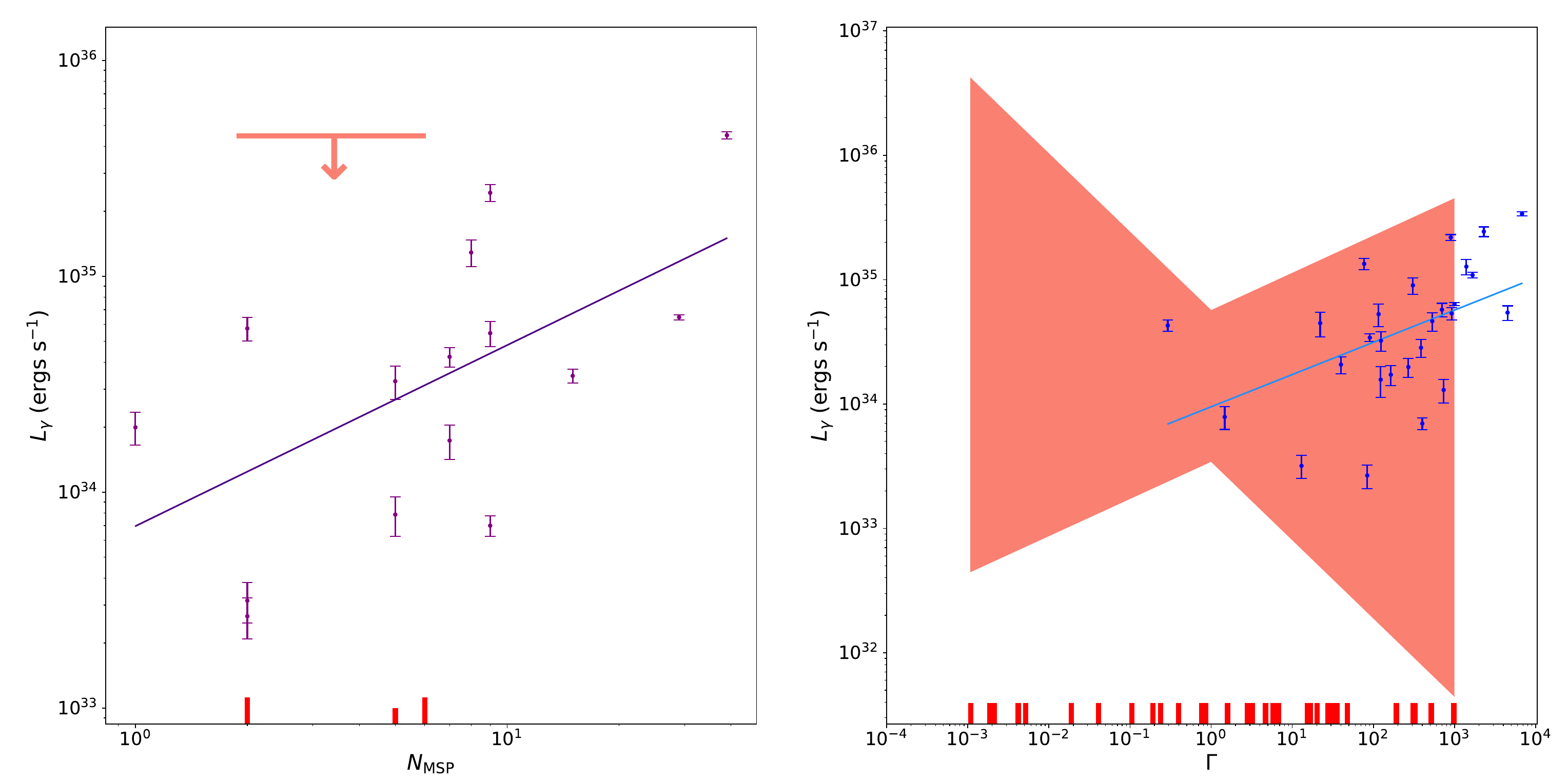}
    \caption{Correlation relations between \lumg\ and \nmsp\ (left) and \enc\ (right) for 4FGL GCs (dots). The red vertical lines on the x-axis represent the values of undetected target GCs. (Left) \lumg\ vs \nmsp: The values of \nmsp\ for the 5 GCs in our target population are taken from \citet{pfreire}. 
    Taller ticks indicate two GCs with that number of MSPs. The salmon upper limit spans the red tick marks on the x-axis. Its height is determined from the $5 \sigma$ contour in Fig.~\ref{fig:param}. The data points are plotted with \nmsp\ from \citet{pfreire} and \lumg\ is computed using the \gray\ energy flux tabulated in the 4FGL and the cluster's distance from \citet{2010arXiv1012.3224H}. (Right) \lumg\ vs. \enc: The error region is centered on the mean \enc\ of the target GCs (Eqn.~\ref{eqn:butt}). The salmon error region is discussed in \S \ref{sec:corr}. The data points are plotted using \enc\ from \citet{2013ApJ...766..136B}, and similarly \lumg\ is calculated from the energy flux in 4FGL and the cluster distance from \citet{2010arXiv1012.3224H}.}
    \label{fig:corr}
\end{figure*}

In the left panel of Fig.~\ref{fig:corr} we plot \lumg\ against \nmsp\ of the detected GCs in the 4FGL with tabulated values of \nmsp\ from \citet{pfreire}. The linear regression of log(\lumg) and log(\nmsp) for the detected 4FGL GCs returns a coefficient of determination of $R^{2} = 0.37$ indicating a weak correlation between the parameters. No upper limits were used in computing this regression. Overlaid on this plot of detected 4FGL GCs we estimate the \lumg\ upper limit of our stacked target GC population by integrating the spectra within the $5\sigma$ contour in Fig. \ref{fig:param}. The maximum energy flux is scaled by $4\pi d^2$, where $d=31.8$ kpc is
the median distance of our targets. This calculation provides the salmon upper limit in the left panel of Fig. \ref{fig:corr}. 

\par In the right panel of Fig.~\ref{fig:corr}, we also perform a linear regression between \lumg\ and \enc\ of detected 4FGL GCs that have a tabulated \enc\ from \citet{2013ApJ...766..136B}. This fit between \lumg\ and \enc\ returns a coefficient of determination of $R^{2} = 0.20$, also indicating a weak correlation. Once again no upper limits were used in computing this regression. For the undetected target GCs, we test for correlation using a technique described in \citet{2023arXiv231019888K} since there are far more target GCs with tabulted \enc\ (34) than there are with \nmsp\ (5). We assume a correlation between \lumg\ and \enc\ in the following form: 
\begin{equation}
    \log{L_{\gamma}} = a+b\log{\Gamma}
    \label{eqn:butt}
\end{equation}
and then explore a grid of slopes ($-2 \leq b \leq 2$) and intercepts ($20 \leq a \leq 40$) for each target GC.

Each \lumg\ is then converted into an energy flux and compared to that individual target's parameter space stack result (Fig.~\ref{fig:param}, described in \S ~\ref{sec:param}) to determine the TS (and thus the likelihood) of that luminosity. From this, we determine the most likely relationship between \lumg\ and \enc\ along with the error region shown in Fig.~\ref{fig:corr} for our target GCs. Our result is consistent with a lack of correlation and also matches the weak correlation between the detected between \lumg\ and \enc\, with a measured power law index of $ b = -0.26_{-0.77}^{+0.56}$.

\begin{figure}
    \centering
     \includegraphics[width = \columnwidth]{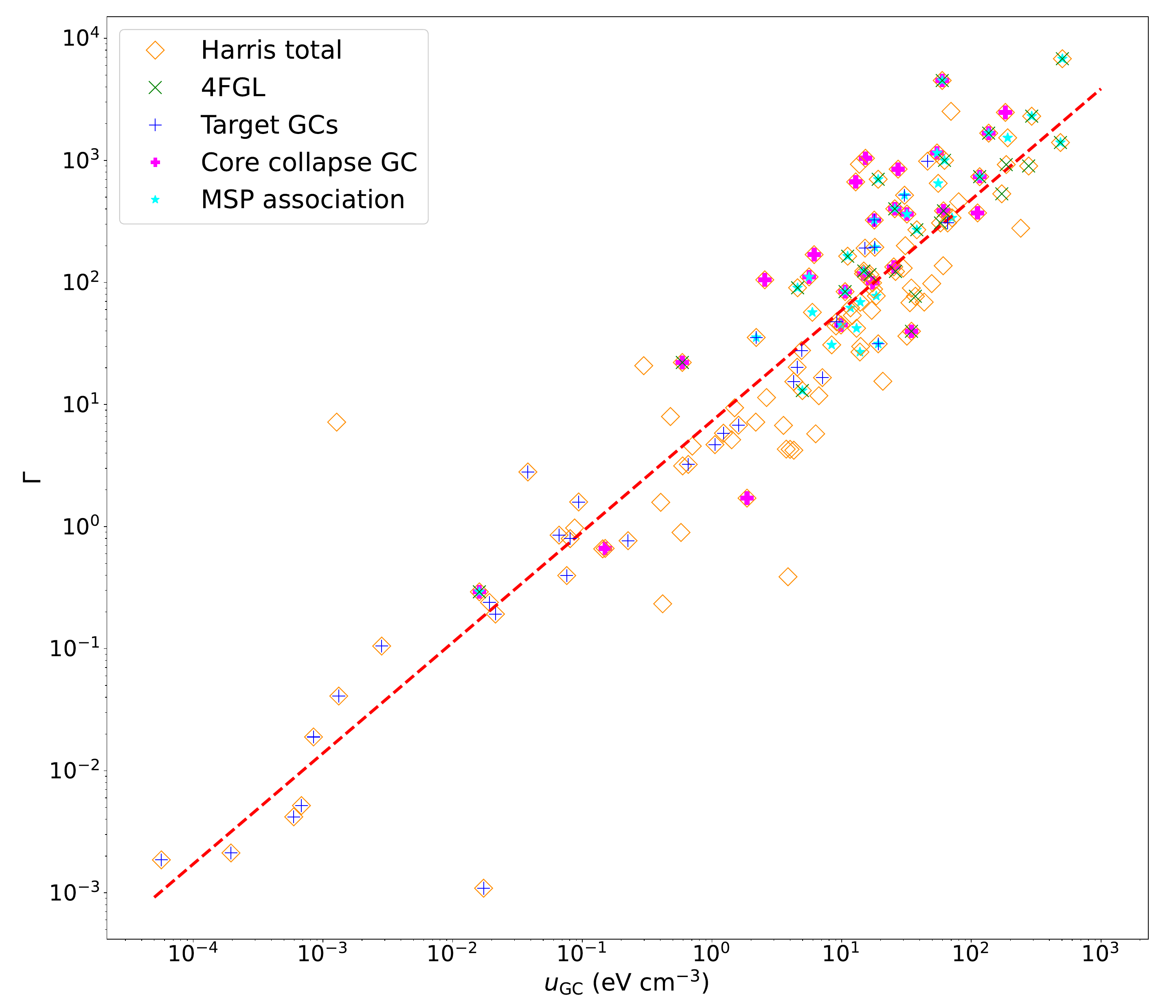}
    \caption{Encounter rate ($\Gamma$) vs. cluster photon field density ($u_{\rm GC}$).  
    Also plotted are different kinds of GCs, such as whether the cluster is core-collapsed, is known to host MSPs, or is in the 4FGL catalog. The red dotted line is the trend from a linear regression (Eqn. \ref{eq:5}).}
    \label{fig:UVG}
\end{figure}

We test the correlation between \enc\ and the photon field density of the cluster, $u_{\rm GC}$, of every GC in \citet{2010arXiv1012.3224H} similar to the ``hidden correlation'' analysis done by \citet{2021MNRAS.507.5161S} (Fig.~\ref{fig:UVG}). The total photon field density has two components: due to the Milky Way ($u_{\rm MW}$) and due to the GC itself ($u_{\rm GC}$), defined as $u_{\rm GC} = L_{*}/{4 \pi c R_h^2}$ \citep{2021MNRAS.507.5161S}. Here $R_h$ is the half-light radius and the stellar luminosity, $L_*$, is estimated from the central luminosity density of the cluster multiplied by the surface integral of the 1-D King model \citep{1962AJ.....67..471K} done in \texttt{AstroPy} \citep{2022ApJ...935..167A}. The cluster parameters are taken from \citet{2010arXiv1012.3224H}. We compute $u_{\rm MW}$ from the ultraviolet - infrared interstellar radiation field model of \citet{2017MNRAS.470.2539P, 2011A&A...527A.109P}. In Fig.~\ref{fig:UVG} we demonstrate the correlation between \enc\ and $u_{\rm GC}$. Detected GCs have both large \enc\ and $u_{\rm GC}$ while our targets have considerably lower values of each. Although this relationship is expected since both quantities depend on the stellar density of the cluster, it is notable that the relation holds over eight orders of magnitude with a coefficient of determination of $R^2 = 0.83$. The fitted trendline is given by
\begin{equation}\label{eq:5}
    \log \Gamma = (0.91 \pm 0.04) \log u_{\rm GC} + (0.86 \pm 0.06).
\end{equation}

Ultimately, we find no strong correlations between cluster properties and \lumg. Additionally, we agree with \citet{2021MNRAS.507.5161S} that the strong underlying correlation between $u_{\rm GC}$ and \enc\ can potentially lead to spurious claims when looking for fundamental planes dependent on three or more variables. 
 
\section{Conclusions}\label{sec:5}

In this work, we study \gray\ emission from 39 previously undetected high-latitude GCs using Fermi-LAT. Our cumulative stacking analyses return a stack separation of $\sim4\sigma$ from CF test sources, which alludes to a significant population of sub-threshold GCs. We find either weak or no significant correlations between \lumg , \enc , and \nmsp\ across this study's test populations (target GCs or detected 4FGL GCs), but recover the strong correlation of \enc\ with photon field density $u_{\rm GC}$ found in \citet{2021MNRAS.507.5161S}.

\section*{Acknowledgements}
This work was supported in part by the National Science Foundation under awards AST-1831412 and AST-2219090, by the NASA New York Space Grant Consortium (NYSG) under award \# 80NSSC20M0096, and a NYSG Student Support Award to O.H. This project made use of computational systems and network services at the American Museum of Natural History supported by the National Science Foundation via Campus Cyberinfrastructure Grant Awards \# 1827153 (CC* Networking Infrastructure: High-Performance Research Data Infrastructure at the American Museum of Natural History). The authors would also like to thank the referee of this publication for thoughtful insight into this study.

\section*{Data Availability}

Fermi-LAT photon data are available through the Fermi-LAT data server\footnote{https://fermi.gsfc.nasa.gov/cgi-bin/ssc/LAT/LATDataQuery.cgi}. Fermi-LAT analysis results (the output .xml or .npy files from Fermipy) of target clusters and control fields can be shared upon request to the first author. Post-processing analysis scripts, including the model to determine the stack divergence significance,  will be shared upon request to the first author, given that a proper citation to this paper is provided in the work from those making the request.



\bibliographystyle{mnras}
\bibliography{ref} 




\appendix

\section{Marginal Detections of NGC~7099 and IC~4499}\label{appxA}

\begin{figure}
    \centering
    \includegraphics[width=\columnwidth]{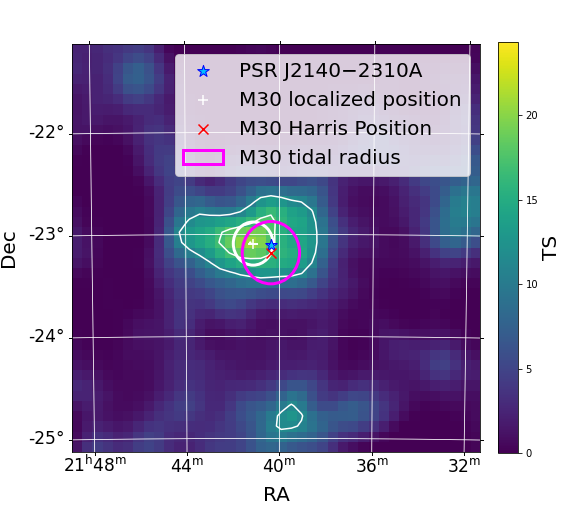}
    \includegraphics[width=\columnwidth]{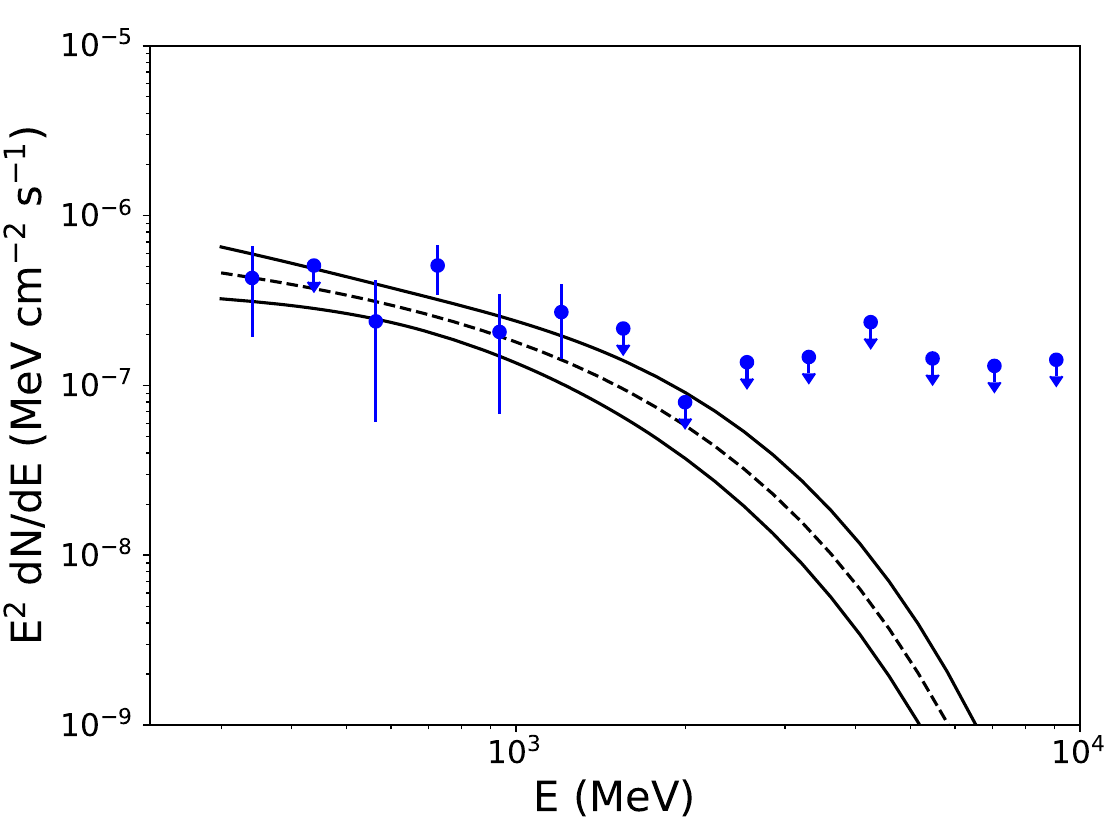}
    \caption{\texttt{fermipy} output data for NGC~7099. (Top) TS map of the central region of the ROI. The tidal radius and catalog position are plotted (magenta and red) from \citet{2010arXiv1012.3224H}. The \texttt{fermipy} localized position is plotted along with its error radius (white) according to the numbers in Appendix \ref{appxA}. The $3 \sigma$ and $5  \sigma$ contours are plotted in white according to the model's two degrees of freedom. The position of the associated pulsar is plotted in cyan from \citet{pfreire}. (Bottom) \texttt{fermipy} computed spectra with a power law index $\gamma = -2.19 \pm 0.44$ and prefactor $N_0 = (4.8 \pm 1.6) \times 10^{-13}$ (ph cm$^{-2} \text{s}^{-1}$). Fluxes with error bars are shown in each energy bin with TS $> 4$. The rest are $95\%$ CL upper limits.}
    \label{fig:M30}
\end{figure}

\begin{figure}
    \centering
    \includegraphics[width=\columnwidth]{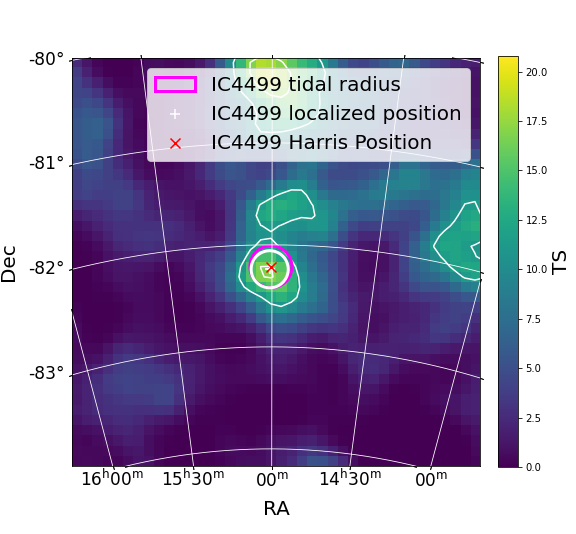}
    \includegraphics[width=\columnwidth]{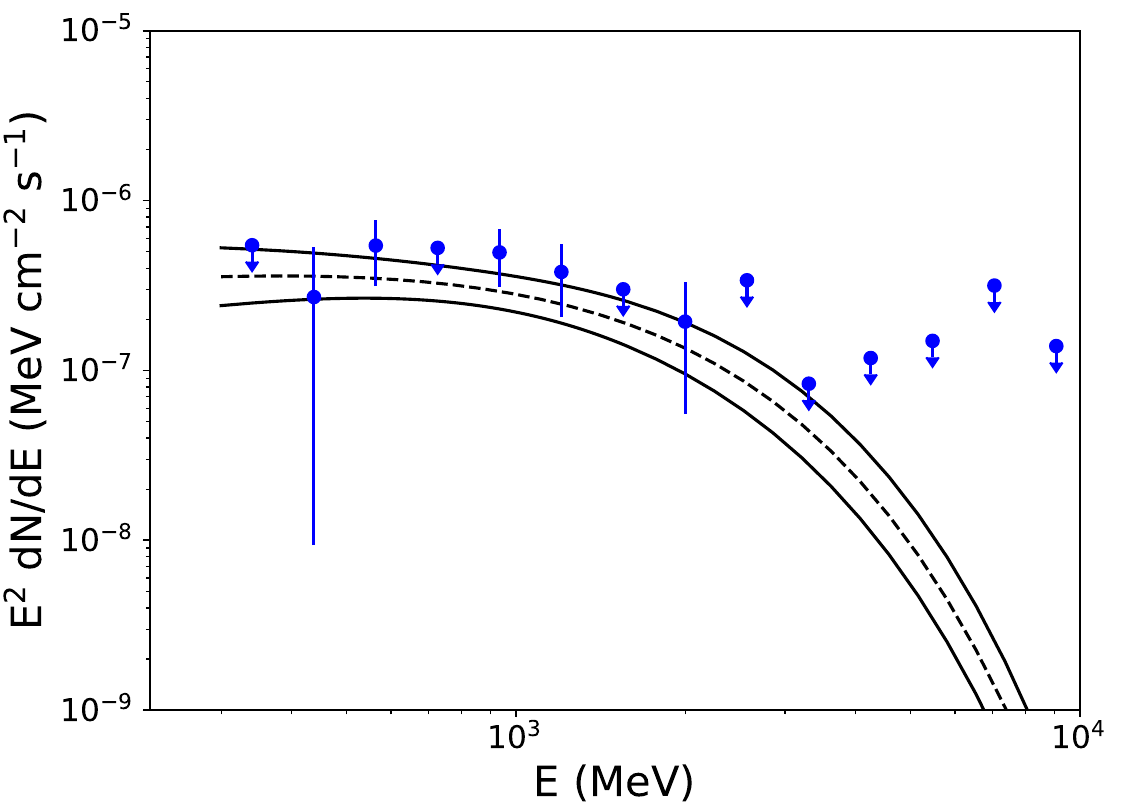}
    \caption{\texttt{fermipy} output data for IC~4499. (Top) TS map of the central region of the ROI. The tidal radius and the catalog position are plotted  (magenta and red) from \citet{2010arXiv1012.3224H}. The \texttt{fermipy} localized position is plotted along with its error radius (white) ccording to the numbers in Appendix \ref{appxA}. The $3 \sigma$ and $5  \sigma$ contours are plotted in white according to the model's two degrees of freedom. (Bottom) \texttt{fermipy} computed spectra with a power law index $\gamma = -1.62 \pm 0.38$ and prefactor $N_0 = (7.6 \pm 2.0) \times 10^{-13} $ (ph cm$^{-2} \text{s}^{-1}$).  Fluxes with error bars are shown in each energy bin with TS $> 4$. The rest are $95\%$ CL upper limits.}
    \label{fig:4499}
\end{figure}

Two sources in our analysis had TS $> 16$, namely NGC~7099 (M~30) and IC~4499 (Table~\ref{tab:table}). 
In Figs.~\ref{fig:M30} and \ref{fig:4499} we plot their TS maps from \texttt{fermipy} along with position data and tidal radius from \citet{2010arXiv1012.3224H}. The \texttt{fermipy} localized positions and their errors are also plotted. In the case of NGC~7099 we have the location of a known MSP that is associated with the cluster \citep{pfreire}. Alongside these TS maps, we show the SED plotted with \texttt{fermipy} according to the spectral analysis described in \S \ref{sec:model}.

We investigate these sources further by finding the peak TS in the sources' respective maps using the \texttt{localize} routine in \texttt{fermipy}. The peak localized TS 
is comparable to the original 
TS value with localized sexagesimal coordinates of ($325.281\pm 0.150$, $-23.090\pm 0.149$) and ($225.165\pm0.133$, $-82.239\pm0.124$) for  NGC~7099 (M~30) and IC~4499 respectively. The original coordinates can be found in Table \ref{tab:table}. The catalogued central cluster location, the \texttt{fermipy} localized position, and $1 \sigma$ error radius all fall within the tidal radius of NGC~7099. For completeness, we also re-optimize the ROI. The combination of having a TS $>16$ and a pulsar detected within it makes NGC~7099 an intriguing source for follow-up with continued Fermi observations and analysis of the LAT photon data to identify any \gray\ pulsations or gather evidence for other non-thermal emission processes. 
In the case of IC~4499, there are no known pulsars within the cluster. IC~4499 is also a relatively low-density, low-encounter rate GC in our sample. The tidal radius, \texttt{fermipy} localized position, and error radius also agree with each other well. So, this marginally detected \gray\ emission points to a need for further radio and \gray\ follow-up observations in search of an emitting source like a pulsar. 

\bsp	
\label{lastpage}
\end{CJK*}
\end{document}